\documentclass{birkjour}

\usepackage[noadjust]{cite}
\usepackage{xcolor}
\RequirePackage[all]{xy}

\usepackage{amssymb} 
\usepackage{amsfonts} 
\usepackage{amsmath} 
\usepackage{slashed} 

\usepackage{egothic}%
\usepackage{pgothic}%
\usepackage{yfonts}%

\usepackage{prodint}

 \usepackage{yhmath} 
  
\theoremstyle{definition}

\theoremstyle{remark}

\newcommand{\BibTeX}{B\kern-0.1emi\kern-0.017emb\kern-0.15em\TeX}
\newcommand{\XYpic}{$\mathrm{X\kern-0.3em\raisebox{-0.18em}{Y}}$-$\mathrm{pic}\,$}

\newcommand{\cl}{C \kern -0.1em \ell}  

\newcommand{\ed}{\end{document}}

 \usepackage{yhmath} 
  
\newcommand{\beq}{\begin{equation}}
\newcommand{\eeq}{\end{equation}}
\newcommand{\beqa}{\begin{eqnarray}}
\newcommand{\eeqa}{\end{eqnarray}}
\newcommand{\nn}{\nonumber}

\newcommand{\pbr}[2]{ \{ \hspace*{-2.6pt} [ #1 , #2\hspace*{1.4 pt} ] 
\hspace*{-2.6pt} \} }

\newcommand{\we}{\wedge}
\newcommand{\der}{\partial}

\newcommand{\inn}{\hspace*{2pt}\raisebox{-1pt}{\rule{6pt}{.3pt}\hspace*
{0pt}\rule{.3pt}{8pt}\hspace*{3pt}}}

\newcommand{\ka}{\varkappa}

\newcommand{\Psib}{\overline{\Psi}}
\newcommand{\Phib}{\overline{\Phi}}

\newcommand{\what}[1]{\widehat{#1}}

\newcommand{\bx}{{\mathbf{x}}}

\newcommand{\BPsi}{{\bf \Psi}} 
   
\DeclareMathOperator{\Tr}{Tr} 

\newcommand{\ugamma}{\underline{\gamma}}

\begin{document}

\title[Milgromian acceleration and the cosmological constant]{\Huge The Milgromian acceleration 
 and \\ 
 the cosmological constant from  \\
precanonical quantum gravity{$^*$}\footnote{$^*$Based on our presentations at MOND40 (St. Andrews, UK), 
XL WGMP (Bialowieza, Poland), EREP 2023 (Bilbao, Spain), D$\Lambda$rk Energy (Frascati, Italy) and POTOR 9th (Krakow, Poland) in June-September 2023.}  }
\author[Kanatchikov]{Igor V. Kanatchikov}
\address{ 
National Quantum Information Centre KCIK, Gdansk, Poland \\
IAS-Archimedes Project, Saint-Rapha\"el, C\^{o}te d’Azur,
France} 
\email{kanattsi@gmail.com}
\author[Kholodnyi]{Valery A. Kholodnyi}
\address{WPI, Vienna, Austria $\&$ Unyxon, Woodforest, TX, USA}
 \email{valery.kholodnyi@gmail.com} 
 
%

\begin{abstract}
We show that the Milgromian acceleration of MOND and the cosmological constant can be understood and quantified as the 
effects of quantum fluctuations of spin connection which are described by precanonical quantum gravity put forward by one of us earlier.   
We also show that a MOND-like modification of Newtonian dynamics at small accelerations emerges from this picture in the non-relativistic approximation. 
\end{abstract}
\maketitle

\tableofcontents

\section{Introduction}

The precanonical field quantization \cite{ik5e,ik2,ik3,ik4,ik5}, which is   
based on the De Donder-Weyl (DDW) Hamiltonian theory known in the calculus of variations 
\cite{rund}, 
has been recently applied to 
general relativity in metric variables \cite{ikm1,ikm2,ikm3,ikm4}, Palatini vielbein gravity \cite{ikv1,ikv2,ikv3,ikv4,ikv5}, and the teleparallel equivalent of general relativity \cite{iktp1,iktp2}. In this formulation, the space-time decomposition and the infinite-dimensional canonical Hamiltonian treatment are not a prerequisite for field quantization. The procedure of precanonical quantization is based on the Dirac 
quantization of the Heisenberg-like subalgebra of Poisson-Gerstenhaber brackets of differential forms 
 that represent dynamical variables
within the DDW Hamiltonian formulation \cite{ik5,mybr1,mybr2,mybr3} and their generalization to the DDW Hamiltonian systems with constraints 
 \cite{mydirac}. 
\nopagebreak 
The result is a hypercomplex generalization of quantum mechanics to field theory where quantum fields are described in terms of 
the wave functions on the bundle of field variables over space-time 
\nopagebreak 
which take values in the Clifford algebra of space-time and satisfy the 
covariant 
\nopagebreak 
analogue of the Schr\"odinger equation whose most general form 
reads 
 \nopagebreak 
\beq  \label{pseq}
(\what{i\hbar\varkappa\slashed \nabla - {H}}) \Psi =0 ,
\eeq
where $\hat{H}$ is the operator 
of the DDW Hamiltonian function 
 defined from 
 the Lagrangian function $L=L(\phi^a, \der_\mu\phi^a, x^\nu)$ as follows: 
\beq  \label{Hp}
H:= \der_\mu {\phi^a} p^\mu_{\phi^a} -L, \quad  p^\mu_{\phi^a} := \frac{\der L}{\der \der_\mu \phi^a}, 
\eeq
$\hat{\slashed \nabla}$ is the covariant Dirac operator on the spacetime 
 (with the Dirac matrices and/or spin connection becoming differential operators in the context of quantum gravity \cite{ikv1,ikv2,ikv3,ikv4,ikv5,iktp1,iktp2}) 
 and the parameter $\varkappa$ is an ultraviolet quantity of the dimension of the inverse spatial volume which 
 appears on purely dimensional grounds. 
The parameter $\varkappa$ also appears in the representations of operators 
 that correspond to differential forms, e.g. 
 the 3-dimensional volume element $d\bx := dx^1\we dx^2\we dx^3$ in 4-dimensional Minkowski spacetime is mapped to 
\beq \label{qmap}
d\bx \mapsto \frac{1}{\ka}\, \ugamma{\, }_0 , 
\eeq
where $\ugamma{\,}^I$ are the flat spacetime Dirac matrices, $\ugamma{\, }^I \ugamma{\, }^J + \ugamma{\, }^J\ugamma{\, }^I 
= 2 \eta^{IJ}$ 
$(I,J = 0,1,2,3)$. 
With $\gamma^\mu := e^\mu_I \ugamma^I$ the operators of polymomenta $p^\mu_{\phi^a}$ defined in (\ref{Hp}) typically have the form 
\beq \label{prep}
\hat{p}{}^\mu_{\phi^a} = -i\hbar\ka \gamma^\mu  \frac{\der}{\der \phi^a} .     
\eeq

Considering the connection of precanonical formulation with standard QFT in the 
Schr\"odinger representation  \cite{iky3,iks1,iks2,iksc1,iksc2,iksc3},  
we conclude that 
 the latter can be 
viewed as a specific limiting case of the former,  
 which corresponds 
to the infinitesimal value of $1/\ka$ or, 
more precisely, to the inverse of the ``Chevalley map" (\ref{qmap}) 
(cf.\cite{meinrenken}). In this case, the Schr\"odinger wave functional $\BPsi ([\phi(\bx)],t)$  can be expressed as a trace of the 
multidimensional Volterra product integral \cite{volterra} of precanonical wave functions $\Psi (\phi,x)$ restricted to the surface 
of initial data $\Sigma$ at the moment of time $t$, $\Sigma\!\!: (\phi=\phi(\bx), \bx, t)$. 
It was also shown that, in this limiting case,  the Schr\"odinger equation for the time evolution of the functional $\BPsi ([\phi(\bx)],t) = 
\BPsi ([\Psi|_\Sigma (\phi(\bx),\bx,t )], [\phi(\bx)])$   follows from the precanonical Schr\"odinger equation (\ref{pseq}) 
 without resorting to the standard procedure of canonical quantization. 
  
In this paper, based on a previous work by one of us, 
we first recall the results of the precanonical quantization of tetrad general relativity in Sect.~2, 
and then, in Sect.~3, discuss the simplest solution of the precanonical Schr\"odinger equation for quantum gravity which corresponds to the quantum analogue of the Minkowski space-time. This allows us to discuss the emergence of the acceleration threshold $a_*$ in Sect.~3, the cosmological constant in Sect.~4, and a modification of the Newtonian dynamics, and its relation to the Milgromian MOND in Sect.~5. The numerical values of the quantities in question are discussed in Sect.~6 and the conclusions are drawn in Sect.~7. 

\section{Precanonical quantum tetrad gravity}

Let us recall the key results of precanonical quantization of tetrad gravity based on the previous work by one of us 
\cite{ikv1,ikv2,ikv3,ikv4}. Starting from the standard Palatini Lagrangian density for general relativity in tetrad variables 
\beq 
{\mathfrak L}= \frac{1}{8\pi G} \left( {\mathfrak e} e^{\alpha}_I e^{\beta}_J \left(\der_{[\alpha} \omega_{\beta ]}^{IJ} +\omega_{[\alpha}^{IK}\omega_{\beta ] K}{}^J \right)  
 - \Lambda {\mathfrak e}\right)  , 
\eeq
where the tetrad coefficients $e^{\alpha}_I$ and the spin connection coefficients 
$\omega_\alpha^{IK}$ are the independent field variables, ${\mathfrak e} := \det(e^I_\mu)$, and by following the procedure of the DDW Hamiltonian formulation, we define the polymomenta of the field variables $e$ and $\omega$, that leads to the primary constraints 
(in the sense of the DDW Hamiltonian formulation), viz.,  
\begin{align}
\begin{split}
{\mathfrak p}{}^\alpha_{e^I_\beta} 
:= \; \frac{\der {\mathfrak L} }{\der\, \der_\alpha e^I_\beta} \; \approx 0, \quad 
 {\mathfrak p}{}^\alpha_{\omega_\beta^{IJ}} &:=\frac{\der {\mathfrak L} }{\der\, \der_\alpha{\omega_\beta^{IJ}}} \approx \frac{1}{8\pi G}
{\mathfrak e} e^{[\alpha}_Ie^{\beta ]}_{J },  
\end{split}
\end{align}
and the DDW Hamiltonian density on the surface of constraints, viz.,  
 \beq
 \mathfrak{e} H := {\mathfrak p}{}_\omega\der \omega + {\mathfrak p}{}_e \der e - {\mathfrak L} 
 \approx -{\mathfrak p}{}^\alpha_{\omega_\beta^{IJ}}\omega_\alpha^{IK}\omega_{\beta K }{}^J 
 + \frac{1}{8\pi G}\Lambda {\mathfrak e} . 
 \eeq
The analysis of constraints according to \cite{mydirac} identifies them as second-class, 
and the calculation of the corresponding generalized Dirac brackets of forms \cite{mydirac} leads to very 
simple expressions such as ($\upsilon_{\alpha} := \der_\alpha \inn dx^0\we dx^1\we dx^2 \we dx^{3}$)
 \beqa 
  \label{dbr11c}
 {}&\pbr{{\mathfrak p}^\alpha_e}{e' \upsilon_{\alpha'}}{\!}^D=0, \nn \\
{}&\pbr{p^\alpha_\omega}{\omega'\upsilon_\beta}{\!}^D
= \delta^\alpha_\beta \delta_\omega^{\omega'}, 
\label{dbr12c} 
 \\
{}&\!\!\hspace*{-10pt}\pbr{{\mathfrak p}^\alpha_e }{ {\mathfrak p}_\omega \upsilon_{\alpha'}}{\!}^D 
\!=\pbr{{\mathfrak p}^\alpha_e }{\omega \upsilon_{\alpha'}}{\!}^D 
\!=\pbr{{\mathfrak p}^\alpha_\omega}{e' \upsilon_{\alpha'}}{\!}^D \!=0 . \,\label{dbr13c} \nn
\eeqa 

Using the generalized Dirac's quantization rule
$[\hat{A}, \hat{B}] = - i\hbar \widehat{\mathfrak{e} \pbr{A}{B}{}^D}$, 
the following representations of relevant operators can be obtained: 
\begin{align}
\hat{e}{}^\beta_I  & = -  8\pi G \hbar\ka i  \ugamma{}^{J}\frac{\der}{\der \omega_{\beta}^{IJ}} ,\label{ebiop} 
\\
\hat{g}{}^{\mu\nu} & = - (8\pi G)^2 \hbar^2\ka^2 \eta^{IK} \eta^{JL} \der_{\omega_\mu^{IJ}} \der_{\omega_\nu^{KL}},   \label{gop} \\
 \what{H} &= 8\pi G \hbar^2\ka^2 \
 \ugamma^{IJ}  
\omega_\alpha^{KM}\omega_{\beta M}{}^L \frac{\der}{\der \omega_{\beta}^{KL}} \frac{\der}{\der \omega_{\alpha}^{IJ}}
+ \frac{1}{8\pi G} \Lambda , \label{hDDWoper}   \\
\what{\not\hspace*{-0.2em}\nabla} &= 
- 8\pi i G \hbar\ka \ugamma{}^{IJ}\frac{\der}{\der \omega_{\mu}^{IJ}} 
\left(\der_\mu +  \frac{1}{4}\, \omega_{\mu KL}\ugamma{}^{KL} \stackrel{\leftrightarrow}{\vee}\right) ,       
\label{nablaoper}
\end{align}
where the spin connection term in (\ref{nablaoper}) acts on the Clifford-valued precanonical wave function $\Psi(\omega,x)$ by 
 the commutator Clifford product:  
${\ugamma}{}^{IJ} \stackrel{\leftrightarrow}{\vee} \Psi 
 : = \frac12 \left[\ \ugamma{}^{IJ}, \Psi\ \right]$. 
 The operators act on Clifford-algebra-valued precanonical wave functions on the configuration bundle of spin connection variables over spacetime variables, $\Psi(\omega,x)$, whose invariant scalar product has the form 
\beq \label{scprod}
\left\langle \Phi | \Psi \right\rangle 
:=  
\Tr\! \int \prod_{\mu, I,J} d \omega_\mu^{IJ} \, \left( \Phib \, \hat{\mathfrak e}{}^{-6} \Psi \right ),
\eeq
where $\Phib:=\ugamma{}^0\Phi^\dagger\ugamma{}^0$    
and 
$\hat{\mathfrak e}{}^{-1} := \what{\det(e^\mu_I)}$ is constructed from 
$\hat{e}{}^\mu_I$ in (\ref{ebiop}).

Using (\ref{hDDWoper}) and (\ref{nablaoper}), the precanonical Schr\"odinger equation 
for quantum gravity, eq. (\ref{pseq}), takes the form 
\beq  \label{psegrav}
\ugamma{}^{IJ}  \frac{\der}{\der \omega_{\mu}^{IJ}}   
   \Big ( \der_\mu + \frac{1}{4} \omega_{\mu}^{KL}\ugamma{}_{KL}\!\! \stackrel{{\leftrightarrow}}{\vee}  
 - \frac{\der}{\der \omega_{\beta}^{KL}}
\omega_{\mu M}^{K}\omega_{\beta}^{ML} \Big)  
 \Psi (\omega,x)     
 -  \lambda \Psi (\omega,x) = 0 ,  
 \eeq
where all the physical constants and the parameter $\ka$ of precanonical quantization are absorbed in the single dimensionless quantity 
\beq \label{lam}
\lambda := \frac{\Lambda}{ (8\pi G \hbar\varkappa)^2} .  
\eeq 

Thus, precanonical quantization leads to the spin connection foam picture of quantum geometry which is described by 
Clifford-algebra-valued amplitudes $\Psi(\omega, x)$ that obey (\ref{psegrav}) and the transition amplitudes 
$\left< \omega, x | \omega', x'\right>$ which are Green's functions of equation (\ref{psegrav}).



  \section{Quantum wave states of Minkowski spacetime}

The Minkowski spacetime in Cartesian coordinates can be characterized by $\omega_\mu^{IJ} = 0$ (cf. \cite{my-mink}).  
In this case, eq. (\ref{psegrav}) with $\Lambda = 0 $ reduces to 
\beq  \label{pseqm}
\ugamma^{IJ}\der_{\omega_\alpha^{IJ}} \der_\alpha \Psi = 0 , 
\eeq
which is solvable by plane waves 
$\Psi \sim e^{ik_\mu x^\mu + i \pi^\alpha_{IJ} \omega_\alpha^{IJ}} \tilde{\Psi} ( \pi^\alpha_{IJ}, k_\mu )$ 
and results in the anisotropic dispersion relation 
 \beq  \label{disp}
 k_\alpha \pi^\alpha_{IJ} k_\beta \pi^\beta{}^{IJ} = 0 .  
 \eeq  
The required correspondence to the Minkowski spacetime on average 
\beq\label{mi9}
\langle \hat{g}^{\mu\nu}\rangle (x) = \Tr\! \int\! d^{24}\omega 
\left( \Psib(\omega,x) 
\hat{\mathfrak{e}}{}^{-6}\hat{g}^{\mu\nu} \Psi (\omega,x) \right) = \eta^{\mu\nu} ,  
\eeq
where the operator 
$\hat{g}^{\mu\nu}$ is given by (\ref{gop}), 
is satisfied by the sufficient condition 
\beq \label{misuf}
\hat{g}^{\mu\nu} \Psi (\omega,x) = \eta^{\mu\nu} \Psi (\omega,x), 
\eeq
which leads to 
\beq \label{mi13}
  \pi_{IJ}^\mu \pi^\nu{}^{IJ}{} = \frac{1}{(8\pi G \hbar\ka)^2} \eta^{\mu\nu} ,  
\eeq 
and, in turn, using the dispersion relation (\ref{disp}), 
to  $k_\mu k^\mu = 0$. 
Therefore, the quantum counterpart of Minkowski spacetime corresponds to the light-like modes of the precanonical wave function  
along the spacetime dimensions,  
and the massive modes along the dimensions of spin connection coefficients.  
The range of 
those modes in the space of spin connection coefficients defines an invariant scale of accelerations 
(in the units in which $c=1$)
\beq \label{astar}
a_* := 8\pi G \hbar\ka .
\eeq 
At this scale, the classical notion of inertial frames is violated by quantum fluctuations of spin connection,  
and the laws of dynamics in external fields can be modified 
 at small accelerations of the order of or smaller than $a_*$. 

\section{The cosmological constant} 

From (\ref{lam}),  
\beq \label{Lala}
{\Lambda} = \lambda{ (8\pi G \hbar\varkappa)^2}, 
\eeq
where the constant $\lambda$ depends on the ordering of operators in (\ref{psegrav}). The ordering is fixed by requiring 
the terms in (\ref{psegrav}) which do not contain the spacetime derivatives $\der_\mu$ to be symmetric operators on the space 
of Clifford-valued wave functions equipped with the scalar product (\ref{scprod}) \cite{kk23}. The contribution from the Weyl ordered 
spin connection operator $\frac{1}{4}  \ugamma{}^{IJ}  \der_{\omega_{\mu}^{IJ}}   
   \omega_{\mu}^{KL}\ugamma{}_{KL}\!\! \stackrel{{\leftrightarrow}}{\vee}$  is 
\beq 
\lambda = -  \frac{1}{16}\,  \ugamma{}^{IJ} \ugamma{}_{KL}  \left[ \der_{\omega_{\mu}^{IJ}} ,   
   \omega_{\mu}^{KL}\right]   = 3 , 
\eeq    
compare with the estimation in \cite{my-mink}. 

\section{Toward MOND from quantum gravity}

Let us consider the non-relativistic motion of a test particle in a gravitational field due to a point mass $M$ and taking into account the fluctuating 
spin connection of Minkowski spacetime. In the non-relativistic limit,  
\beq  \label{mo3}
\ddot{x}{}^i =
- \Gamma^i_{00}= 
- \omega^{i 0}_0 = - GM \frac{x^i}{r^3} - \tilde{\omega}{}^i  , \quad \langle \tilde{\omega}{}^i \rangle = 0 . 
\eeq
 The fluctuations of spin connections $\tilde{\omega}{}^i$ are distributed according to the wave function 
$\Psi(\omega, x)$ that obeys the non-relativistic approximation of (\ref{misuf}):
\beq\label{mo20}
\eta^{ij} \der_{\tilde{\omega}{}^i} \der_{\tilde{\omega}{}^j} \Psi 
= - \frac{1}{(8\pi G \hbar\ka)^2} \Psi ,   
\eeq
whose real-valued ground state (Yukawa) solution and its normalization are 
\beq \label{mo22}
\Psi = 
\frac{1}{\pi \sqrt{8 G \hbar\ka}\ \omega} 
\ e^{-\omega/(8\pi G \hbar\ka) } , 
\quad \omega : = \sqrt{\tilde{\omega}{}^i \tilde{\omega}{}_i} , 
\eeq 
\beq
\langle \Psi |\Psi \rangle = \int\! d {\tilde{\omega}{}^1} d {\tilde{\omega}{}^2} d{\tilde{\omega}{}^3} \, \Psi \Psi = 1 . 
\eeq
By averaging the square of 
(\ref{mo3})
by using $\langle \tilde{\omega}{}^i \rangle = 0$ and 
\beq
 \langle \tilde{\omega}{}^i \tilde{\omega}{}_i \rangle 
 := \bar{a}{}^2 
= \int\! d {\tilde{\omega}{}^1} d {\tilde{\omega}{}^2} d{\tilde{\omega}{}^3} \, 
\Psi \omega^2 \Psi  = \frac{1}{2} (8\pi G \hbar\ka)^2 = \frac{1}{2} a_*^2 , 
\eeq
and, by denoting $\langle \ddot{x}{}^i\ddot{x}_i \rangle 
=: a^2$,
we obtain 
a modification of Newton's law 
 due to quantum 
fluctuations of spin connection: 
 \beq \label{mo9}
 \frac{GM}{r^2} = \sqrt{a^2 + \bar{a}{}^2 }{}  .  
\eeq 
This formula is valid in the region $r\,<\,\sqrt{{GM}/{\bar{a}}}$ where $a^2>0$. At larger distances, the fluctuations of spin connection which 
lead both to the acceleration threshold (\ref{astar}) and the cosmological constant (\ref{Lala}) dominate and hence violate the approximation of global Minkowski spacetime on average. 

When $a\gg\bar{a}{}$, we obtain from (\ref{mo9}) a corrected Newton's law
\beq\label{mo10}
a + \frac{\bar{a}{}^2}{2a} = \frac{GM}{r^2}  \quad \mathrm{for}\; {r} < \sqrt{\frac{2GM}{3\bar{a}}}.  
\eeq
When $\bar{a}{}\gg a$, we obtain  a MOND-like relation (cf. \cite{mond}) 
\beq \label{mo12}
\bar{a}{} + \frac{a^2}{2\bar{a}{}} = \frac{GM}{r^2}  
\quad \mathrm{for}\;  \sqrt{\frac{2GM}{3\bar{a}}}
\,<\,r\,<\,\sqrt{\frac{GM}{\bar{a}}}    
\eeq
if the Milgromian acceleration $a_0$ is identified with $2\bar{a}$. In this case, 
\beq \label{a0lambda}
 a_0 = 2\bar{a} = 8  \sqrt{2} \pi G \hbar\ka  
 = \sqrt{2 \Lambda /  \lambda} .  
\eeq

Note that our consideration here neglects the influence of the fluctuations of spin connection on the central mass $M$ and 
quantum correlations of spin connections at the locations of the mass $M$ and the test particle.
We 
expect that by taking those effects into account we can obtain the Milgromian MOND together with a realistic interpolating 
function $\mu (a/{a_0})$ and 
 relate the Milgromian $a_0$ with theoretically predicted scales $a_*$ and $\bar{a}$ 
more precisely (cf.\cite{kk23}). This may lead to 
a more realistic coefficient 
 in the relation between $a_0$ and $\sqrt{\Lambda}$ than the one in (\ref{a0lambda}) 
(cf. Milgrom's $2\pi a_0\!=\!\sqrt{\Lambda/3}$ in\,\cite{mond}) 
and a more realistic range 
of MOND-like dynamics than in (\ref{mo12}) (cf. Sect. 6).


\section{Numerical estimates}

The numerical values of $a_*$ and $\Lambda$ depend on the value of the parameter $\ka$ introduced by precanonical quantization. 
It is shown in \cite{my-ymmg} that the mass gap of quantum Yang-Mills gauge theory is related to the scale of $\ka$:  
$\Delta m\! \sim\! (g^2\hbar^4\ka)^{1/3}$.  
 In QCD, $\Delta m \sim 10^{0\pm 1} \mathrm{GeV}$  
and $g^2 \sim 10^0$ (see \cite{iktp2} and the references therein), and, therefore, by taking into account all the current 
uncertainties in the values of $\Delta m$, the gauge coupling $g$, and the spectral estimate in \cite{my-ymmg}, 
we conclude that $\ka \sim 10^{0 \pm 6} \mathrm{GeV}^3$. Consequently, from (\ref{astar}) and (\ref{Lala}), we obtain  
\beq
a_* \sim 10^{-23\pm 6} \mathrm{cm}^{-1} \quad \mathrm{and} \quad \Lambda \sim  10^{-46 \pm 12} \mathrm{cm}^{-2} .  
\eeq 
These values overlap with the values of the Milgromian acceleration 
 $a_0 \approx 10^{-29} \mathrm{cm}^{-1}$   
and the cosmological constant 
${\Lambda} \approx 10^{-56} \mathrm{cm}^{-2}$, respectively. 

Correspondingly, 
the corrected Newton's law (\ref{mo10}) is valid in the Solar System up to 
$\sqrt{4 GM_\odot /3 a_0} \sim   6 \times 10^3$ au, i.e. 
 the inner edge of the 
 \"Opik-Oort cloud. 
For a galaxy of total mass $\mathfrak{M} \sim 10^{11} M_\odot$, 
the MOND-like dynamics in (\ref{mo12}) 
is valid 
only 
in a narrow range of 
galactocentric distances between $\sqrt{{4 G\mathfrak{M} }/{3{a_0}}} \sim 11$ kpc 
and $\sqrt{{2 G \mathfrak{M}}/{{a_0}}} \sim 13$ kpc.   
  The agreement with the range of flat rotation curves of galaxies 
  that are described by MOND \cite{mond5} 
  may be improved by taking into account the effects listed in the end of Sect.\,5. 
  \section{Conclusion} 

We found two 
manifestations of quantum fluctuations of spin connection related to the Milgromian acceleration 
$a_0 = \sqrt{2} a_* = 2 \bar{a}$: the range $a_*$ in the spin connection space of precanonical wave function 
 that corresponds 
 to the quantum analogue of Minkowski spacetime, and the standard deviation  $\bar{a}$ of the distribution of spin connections given by the precanonical wave function in the non-relativistic approximation. The cosmological constant also appears as a manifestation of the quantum dynamics of spin connection 
in the form of the 
 appropriate ordering of the spin connection operator in precanonical Schr\"odinger equation (\ref{psegrav}). 
The relation between the cosmological constant and the Milgromian acceleration: 
$ {a_0} \sim \sqrt{\Lambda}$ appears as an elementary consequence of precanonical quantum gravity. The numerical values of $a_0$ and $\Lambda$ 
can be obtained if the parameter $\varkappa$ has a hadronic scale, which is consistent with the estimation of the gap in the spectrum of the DDW Hamiltonian operator of pure gauge theory in \cite{my-ymmg}. It is also shown that quantum fluctuations of spin connection lead to a 
modification of Newtonian dynamics similar to MOND in the regime of very weak gravitational fields. 
Given the phenomenological success of MOND \cite{mond5}  
and the relation of the Milgromian acceleration $a_0$ to the cosmological constant $\Lambda$  and  the Hubble constant $H$ \cite{mond6}, 
as well as the natural appearance of the Milgromian acceleration and the cosmological constant 
 in precanonical quantum gravity and its ability to obtain realistic values of $a_0$ and $\Lambda$, albeit with the current error of several orders of magnitude, we believe that the current trend of introducing {\em ad hoc} entities as 
various candidates for dark matter, dark energy, and modified theories of gravity in the context of galactic dynamics and cosmology 
is worth reconsidering. 

\subsection*{Acknowledgment}
We thank Ilya Kholodnyi for his help with editing the English of the paper.


\end{document}